\documentclass{article}
\usepackage{spconf,amsmath,graphicx}
\usepackage{amsfonts}
\usepackage{amsmath}
\usepackage{subfigure}
\usepackage{mathrsfs}
\usepackage{multirow}
\usepackage{bm}
\usepackage{verbatim}
\usepackage[normalem]{ulem}
\usepackage{float}

\usepackage{mathtools,lipsum,cuted}
\usepackage{xfrac}
\usepackage[colorlinks]{hyperref}
\usepackage[short]{optidef}
\usepackage{bbm}

\usepackage{tikz}
\usepackage{float}

\newsavebox{\tempbox}

\allowdisplaybreaks
\usepackage[]{caption} 
\usepackage{graphicx}
\usepackage[export]{adjustbox}
\usepackage{caption}

\usepackage{enumitem}

\usepackage[ruled]{algorithm}
\usepackage{algpseudocode,algorithmicx}

\newcommand{\vx}{{\boldsymbol x}}

\newcommand{\vE}{{\boldsymbol E}}

\newcommand{\vr}{{\boldsymbol r}}

\newcommand{\vh}{{\boldsymbol h}}
\newcommand{\vg}{{\boldsymbol g}}

\newcommand{\vo}{{\boldsymbol o}}

\newcommand{\mz}[1]{\textsf{\textbf{\color{blue}{\tiny [MZ: #1]}}}}

\title{3D unknown view tomography via rotation invariants}
\name{Mona Zehni$^*$, Shuai Huang$^*$\thanks{$^*$ Equal contribution}, Ivan Dokmani\'{c}, Zhizhen Zhao\thanks{This work is partially supported by NSF CIF-1817577 and NSF DMS-1854791.}}
\address{Department of ECE and CSL, University of Illinois at Urbana-Champaign}

\begin{document}
\savebox{\tempbox}{\begin{tabular}{@{}r@{}l@{\space}}
&\scriptsize{SNR}\\ \scriptsize{$M$}
\end{tabular}}

\newcommand{\remove}[1]{}

\ninept
\maketitle

\begin{abstract}
In this paper, we study the problem of reconstructing a 3D point source model from a set of 2D projections at unknown view angles. Our method obviates the need to recover the projection angles by extracting a set of rotation-invariant features from the noisy projection data. From the features, we reconstruct the density map through a constrained nonconvex optimization. We show that the features have geometric interpretations in the form of radial and pairwise distances of the model. We further perform an ablation study to examine the effect of various parameters on the quality of the estimated features from the projection data. Our results showcase the potential of the proposed method in reconstructing point source models in various noise regimes. 

\end{abstract}

\section{Introduction}
\vspace{-0.1cm}
In 3D unknown view tomography the task is to reconstruct a 3D map from a large set of 2D noisy projections taken from unknown view angles. This paradigm appears in a multitude of applications including cryo-electron microscopy and medical imaging \cite{FRANK19961,van2000}. In this paper we address this problem for a specific form of 3D maps, point-source models. Point-source models are a superposition of a finite number of translated kernels that are well concatenated in space. Recovering a point-source model also appears in a variety of signal and image processing problems, such as compressed sensing \cite{Boche2015}, super-resolution \cite{Candes2012}, radio astronomy \cite{Hanjie2017,Leap2017}, array signal processing \cite{Krim1996}, unassigned distance geometry \cite{Billinge2016,Duxbury2016}, molecular imaging in X-ray crystallography \cite{Drenth2007}, atomic modeling in cryo-electron microscopy \cite{Scheres2012}, powder diffraction \cite{Fabian2006}, to name a few.  

Reconstructing a 3D structure from a set of projection images has been extensively studied in the literature. Techniques targeting this problem can be broadly classified into two categories. In the first category, the projection orientations or their distribution alongside the 3D structure are recovered. One set of such approaches first estimates the projection orientations through common-line based methods~\cite{Penczek1996}, and then recovers the 3D structure through direct filtered backprojection based~\cite{Radermacher2006} or regularized optimization-based methods~\cite{Nilchian2013, donati2018fast}. On the other hand, projection matching~\cite{Barnett2016} and maximum-likelihood based
methods~\cite{Scheres2012, Punjani2017} iteratively estimate the projection orientations and the 3D structure. These conventional methods have major drawbacks such as, 1) they rely on estimating the projection orientations which is a challenging task especially in severe low signal-to-noise ratio (SNR) regimes, 2) they are computationally demanding.

In the second category that is mainly specialized for \textit{ab initio} modeling, recovering the projection orientations is bypassed through the use of rotation-invariant features in an autocorrelation form known as method of moments~\cite{KAM1980, Kam1985, Eithan2017, Sharon2019}. Although this method avoids the estimation of the projection orientations, it does not address how the prior of the signal model could be incorporated in the reconstruction process.

\begin{figure*}[t!]
    \centering
    \includegraphics[width = 0.9 \linewidth]{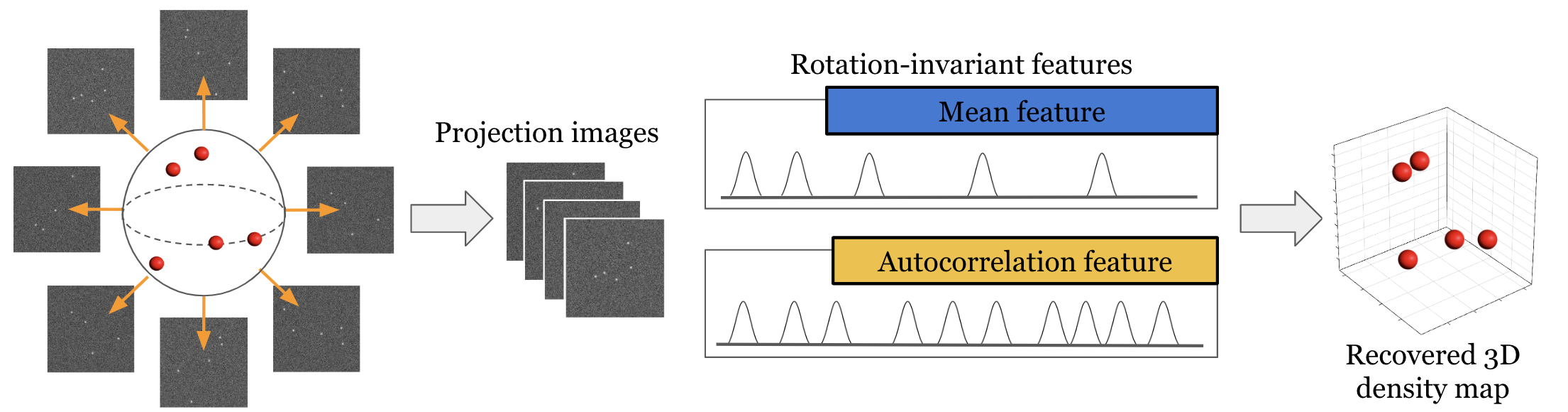}
    \caption{The forward model in~\eqref{eq:forward_model} and the 3D point source localization pipeline. The rotation-invariant features are estimated from the projection images (section 3.1). Finally, the point source model is reconstructed from the features (section 3.2).}
    \label{fig:pipeline}
    \vspace{-0.5cm}
\end{figure*}

In this paper we propose a two-step procedure to recover a 3D point-source model directly from a set of projection images taken at random unknown orientations. As illustrated in Fig.~\ref{fig:pipeline}, our method consists of: (1) estimating rotation-invariant features from the projection data; and (2) reconstructing the density map using the estimated features. We extend our previous work in the 2D point-source tomography~\cite{GeoInv19}-\cite{DistRet19} to 3D, construct new rotational invariant features from the  projection data, and derive analytically the link between the features and the radial and pairwise distances of the points. The features and the derivations are different from our previous results for 2D tomography. Through the use of rotation-invariant features, we bypass the recovery of the projection angles. Compared to other related works on using method of moments to reconstruct 3D density maps~\cite{Eithan2017,Sharon2019}, our approach explicitly takes the prior of the signal, i.e. point-source model, into account in both feature generation and reconstruction step. In the second step of our pipeline, we recover the density map from the estimated features by solving a constrained nonconvex problem using the approach in \cite{Huang2018}. 

We assess the quality of the estimated features through an ablation study and compare the estimated features with their analytical forms. Numerical experiments show that the proposed pipeline is robust to noise when the number of random projections is sufficiently large.

\section{Image formation model}
\vspace{-0.1cm}

\label{sec:forward-model}

We assume the following forward model,
\begin{align}
    s_\ell [u, v] &= \mathcal{D}  \{ \mathcal{P}_{\omega_\ell} \phi \}[u, v] + \varepsilon_\ell [u,v], \quad \ell \in \{1,2,...,L\} \label{eq:forward_model} \\
    \phi(\mathbf{x}) &= \textstyle\sum_{n=1}^{K} \psi(\mathbf{x}-\mathbf{x}_n)\,,
\end{align}
where $\phi:\mathbb{R}^3 \rightarrow \mathbb{R}^+ \cup \{0\}$ is an unknown density map. We further assume that $\phi$ is a point source model consisting of $K$ 
point sources located at $\{\mathbf{x}_n\}_{n=1}^{N_x}$ where $\mathbf{x}$ represents a point in the Cartesian coordinates. The kernel $\psi$ is well concentrated in space, for example, a Gaussian kernel (source). Here, inspired by the atomic modeling for proteins using cryo-electron microscopy images, we assume a simplified model with point sources.


The operator $\mathcal{P}_{\omega}$ projects the 3D density map $\phi$ by first rotating the volume with a $3\times 3$ rotation matrix $\mathcal{R}_\omega$ corresponding to the rotation $\omega$ in 3D rotation group $\mathrm{SO(3)}$, i.e. $ \omega \in \mathrm{SO}(3)$, 
and then taking the line integral of the rotated density map along the $z$-direction,
\begin{align}
\left(\mathcal{P}_{\omega} \phi \right) (x,y) = \int_{-\infty}^{\infty} \phi(\mathcal{R}_{\omega}^\top \mathbf{x}) dz.
\end{align}

To take into account the finite resolution of the digitized projection data, we introduce the sampling operator $\mathcal{D}$ as,
\begin{align}
\label{eq:discretization_image}
\mathcal{D}(f)[u, v] = \int_{\left(v - \frac{1}{2}\right)\Delta}^{\left(v + \frac{1}{2}\right)\Delta} \int_{\left(u - \frac{1}{2}\right)\Delta}^{\left(u + \frac{1}{2}\right)\Delta} f(x, y) dy dx,
\end{align}
where $[u,v] \in \{-M, \dots, M\}\times\{-M, \dots, M\}$, and the pixel width is $\Delta$. The observed discretized projection data is further contaminated by additive white Gaussian noise $\varepsilon$ with zero mean and variance $\sigma^2$. 
The inverse problem we would like to address is to estimate the point source locations in 3D from the collection of noisy projection images $\{s_\ell\}_{\ell = 1}^L$. We emphasize that the rotations $\{{\omega_\ell} \}_{\ell=1}^{L}$ of the projection images are unknown, and we assume $\omega$ is uniformly distributed on $\mathrm{SO}(3)$. 

\section{Method}
\vspace{-0.1cm}
\label{sec:method}
\begin{figure*}
\centering     
\subfigure[$\log_{10} \textrm{SNR} = -0.6, \Delta = 0.005, (N_k, N_\varphi) = (400,400)$]{\label{fig:proj}\includegraphics[width=82mm]{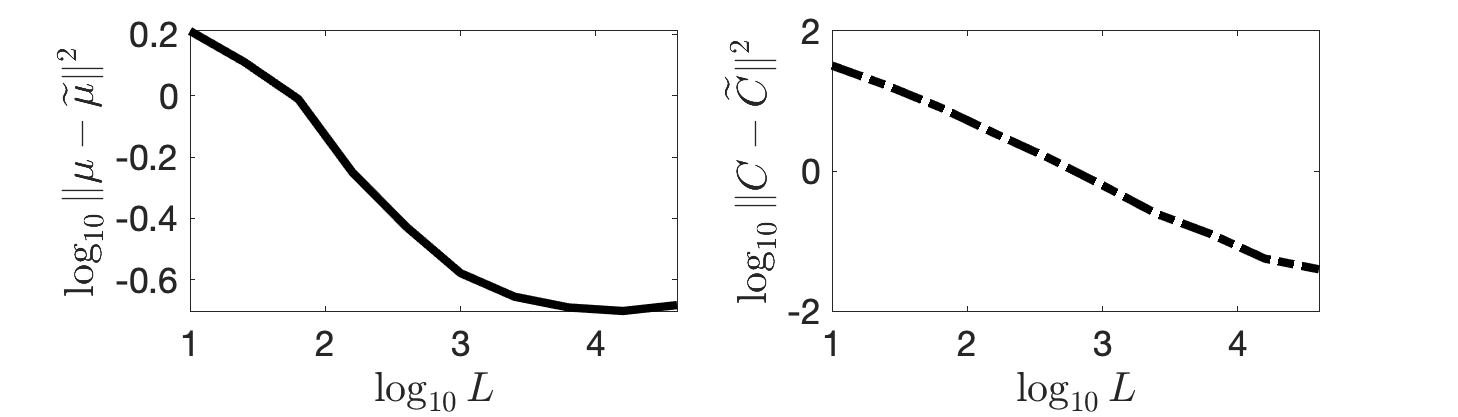}} \quad
\subfigure[$\textrm{SNR}=\infty, L = 10^4, (N_k, N_\varphi) = (400,400)$]{\label{fig:pixel}\includegraphics[width=82mm]{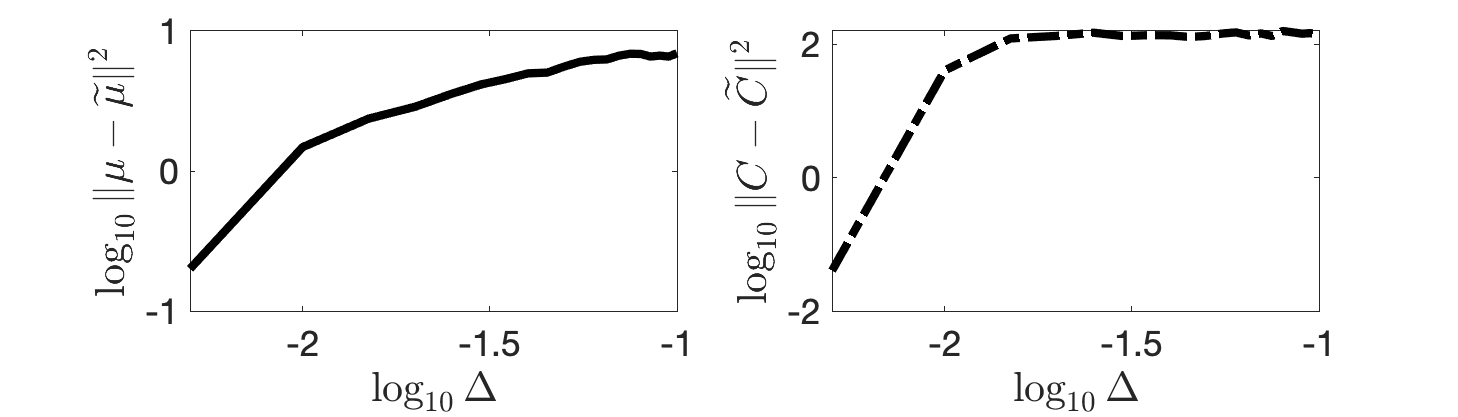}}
\subfigure[$\log_{10} \textrm{SNR} = -0.6, \Delta = 0.005,  L = 2\times 10^4, N_k = N_{\varphi}$]{\label{fig:fp}\includegraphics[width=82mm]{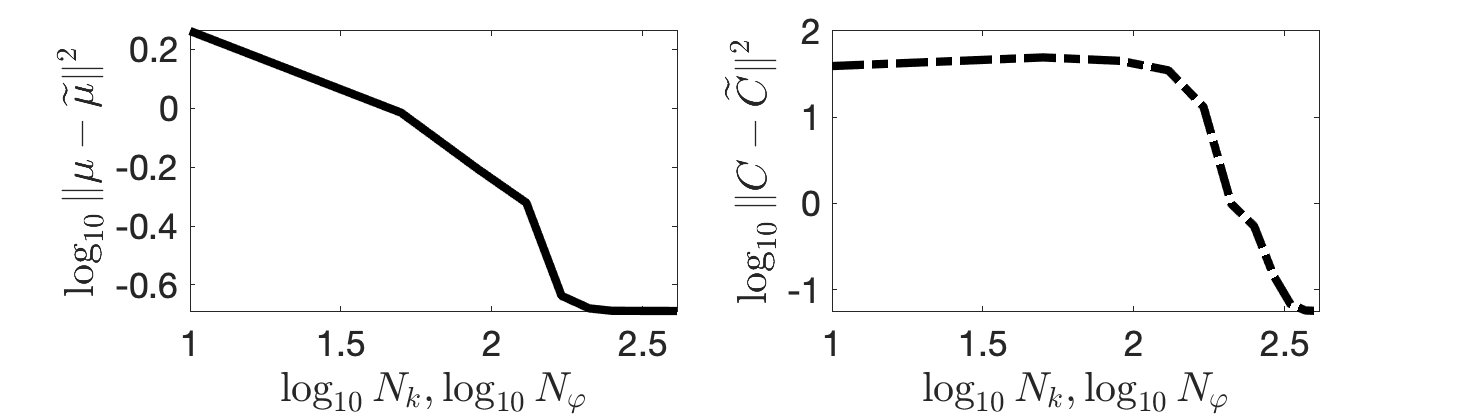}} \quad
\subfigure[$ L=2\times 10^4, \Delta = 0.005, (N_k, N_\varphi) = (400,400)$]{\label{fig:sigma}\includegraphics[width=82mm]{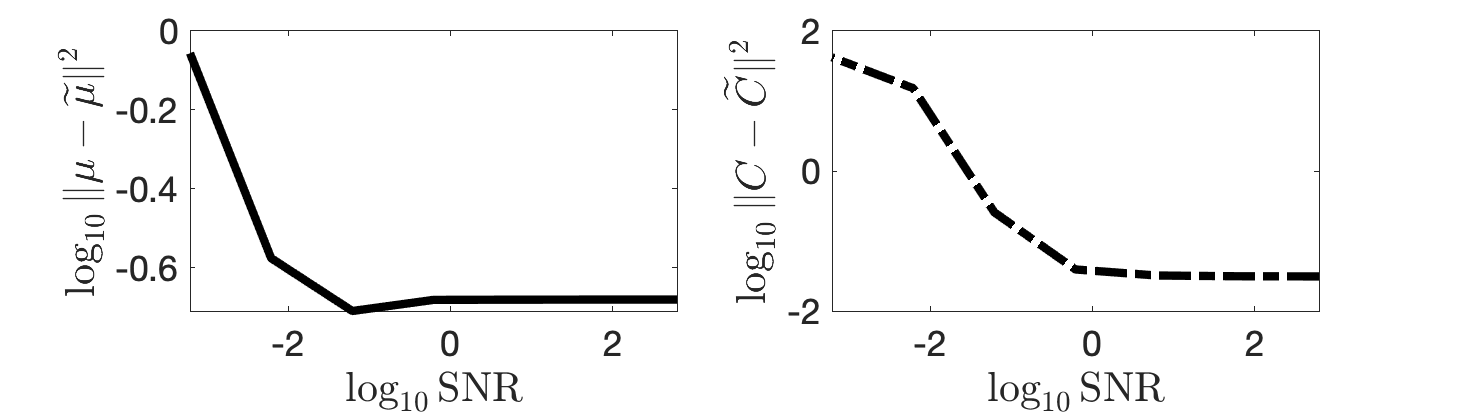}}
\vspace{-0.2cm}
\caption{The ablation study to examine the effect of various parameters on the quality of the estimated mean (solid curves) and autocorrelation features (dashed curves). The parameters under consideration are, \protect \subref{fig:proj} $L$: number of projections, \protect \subref{fig:pixel} $\Delta$: the sampling step, \protect \subref{fig:fp}  $(N_k, N_{\varphi})$: the number of radial and angular points in the polar Fourier grid, \protect \subref{fig:sigma} $\mathrm{SNR}$ of the projection images.}
\label{fig:estimate_features}
\vspace{-0.5cm}
\end{figure*}

We start by deriving the rotation invariant features extracted from the projection data for a ``general'' 3D density map $\phi$, then narrow down to the features for a Gaussian-source model that is later on used in the numerical experiments. Finally, we formulate the reconstruction as a constrained nonconvex optimization problem. 

\subsection{Rotation invariant features}
Let the Fourier transform $\Phi$ of the density map $\phi$ be defined as,
\begin{align}
\Phi(\bm{k}) = \iiint \phi(\bm{r}) e^{-j \langle\bm{k}, \bm{r}\rangle} d\bm{r}, 
\end{align}
where $\bm{r}$ and $\bm{k}$ denote 
coordinates in spatial and Fourier domains respectively with $\Vert \bm{r} \Vert_2 = r$ and $\Vert \bm{k} \Vert_2 = k$. 
From now on, any triple integral with respect to $\bm{r}$ is written as a single integral for the sake of brevity. To obtain the first invariant feature, we average $\Phi(\bm{k})$ over all angular directions of $\bm{k}$ to obtain $B_1(k)$,
\begin{align}
B_1(k) = \int \frac{\sin(kr)}{kr} \phi(\bm{r}) d\bm{r}.
\label{eq:b1_definition}
\end{align}
Note that after averaging, $B_1(k)$ is only a function of $k$. 
Taking the sine transform of $kB_1(k)$ and then multiplying it by $t$, we have
\begin{align}
    \mu(t) &= \frac{2t}{\pi} \int_{0}^{\infty} k B_1(k) \sin(k \, t)\ dk=\int\phi(\vr)\delta(t-r)\ d\vr\,, \label{eq:mu_definition}
\end{align}
for $t\geq 0$.
In~\eqref{eq:mu_definition}, $\mu(t)$ integrates $\phi(\vr)$ on a sphere with radius $t$, and it is thus invariant to the global rotation of the density map around origin. We call $\mu(t)$ the \textit{mean feature} as it is the result of averaging $\Phi(\bm{k})$ over all possible directions of $\bm{k}$. 

For the second feature, we average $\vert \Phi(\bm{k}) \vert^2$ over all angular directions of $\bm{k}$ to get, 
\begin{align}
& B_2(k) 
= \int \frac{\sin(k\|\vr_1-\vr_2\|_2)}{k\|\vr_1-\vr_2\|_2} \phi(\vr_1)\phi(\vr_2)\ d\vr_1d\vr_2.
\label{eq:B2_definition}
\end{align}
We then apply the sine transform to $kB_2(k)$ to get the \textit{autocorrelation feature},
\begin{align}
C(t)= \frac{2t}{\pi}\int_0^\infty kB_2(k)\sin(kt)\, dk
=\int A(\bm{h})\delta(t-\Vert \vh\Vert_2)\ d\vh, 
\label{eq:c_definition}
\end{align}
where $A(\vh)$ is the autocorrelation of $\phi(\vr)$, $ A(\vh)=\int\phi(\vr_1)\phi(\vr_1-\vh)\ d\vr_1 $. As $C(t)$ integrates the autocorrelation function ${A}(\vh)$ on the sphere with radius $t$ (hence the name of the feature), it is also rotation-invariant.

Note that we are calling $\mu$ and $C$ as features as they are functions (or features) of the density map $\phi$. 
Equations ~\eqref{eq:b1_definition}-\eqref{eq:c_definition} reveal how the mean and autocorrelation features are linked to the density map $\phi$. However, since $\phi$ is unknown, we cannot directly compute the features according to \eqref{eq:b1_definition}-\eqref{eq:c_definition}. Thus, our task is to estimate them from the projection data, as described in the following.

The Fourier slice theorem states that the 2D Fourier transform of the projection of a 3D density map taken along direction $\omega$ is exactly the slice of the 3D Fourier transform of the density map perpendicular to $\omega$~\cite{Hsieh2003}. Consequently, the set of all the 3D radial lines of $\Phi(\bm{k})$ is the same as the union of all the 2D radial lines of the projection images taken from all possible angles.  
Consequently, the set containing $\Phi(\bm{k})$ along all 3D radial lines can be obtained by collecting the Fourier transform of all the projection images along the 2D radial lines. Hence, we can write $B_1(k)$ in terms of the projection images 
\begin{align}
B_1(k) = \frac{1}{2 \pi} \int\limits_{SO(3)} \int\limits_{0}^{2 \pi} \left( \mathcal{F} \mathcal{P}_{\omega} \phi \right)(k, \varphi) d \varphi d \omega,  \label{eq:B1_approx}
\end{align}

As we only have access to $L$ projection images, the sample estimate of the feature $B_1$ is,
\begin{align}
\widetilde{B}_1(k_i) \approx \frac{1}{LN_\varphi}\sum\limits_{\ell=1}^{L} \sum\limits_{p=1}^{N_\varphi} \widehat{s}_{\ell}(k_i, \varphi_p), 
\label{eq:B1_estimation}
\end{align}
where $\widehat{s}_{\ell}(k, \varphi)$ denotes the polar Fourier transform of the projection image $s_{\ell}$ at a given point $(k, \varphi)$. Also, $\{(k_i, \varphi_p)\}$ for $i \in \{1,...,N_k\}$ and $p \in \{1,...,N_{\varphi}\}$ denote a set of non-uniformly spaced points in the 2D Fourier space. We use NFFT package~\cite{Keiner2009} to compute the DFT of the projection images sampled on a non-uniformly spaced grid in Fourier domain. Finally, the \emph{mean feature} $\mu(t)$ is approximated by,
\begin{align}
\widetilde{\mu}(t_j) \approx \frac{2 t_j}{\pi}\sum_{i = 1}^{N_k} w(k_i) k_i \widetilde{B}_1(k_i) \sin(k_i \, t_j), \label{eq:mu_approx}
\end{align}
where the upper limit of the integral in~\eqref{eq:mu_definition} is replaced with a cutoff frequency $c$, $t$ is discretized to the finite set $\{t_j\}_{j=1}^T$ and the integral is computed using the Gauss-Legendre quadrature rule~\cite[Chap. 4]{press1992numerical} on $N_k$ points in the interval $[0, c]$ with the associated weights $w(k_i)$.

Following the same steps, the finite sample estimate of the \emph{autocorrelation feature} $C(t)$ from the projection images is,
\begin{align}
\widetilde{B}_2(k_i) &\approx \frac{1}{L N_{\varphi}} \sum\limits_{\ell=1}^{L} \sum\limits_{p=1}^{N_{\varphi}} \vert \widehat{s}_{\ell}(k_i, \varphi_p) \vert^2 \\ 
\widetilde{C}(t_j) &\approx \frac{2t_j}{\pi} \sum_{i = 1}^{N_k} w(k_i) k_i \widetilde{B}_2(k_i) \sin(k_i \, t_j).
\label{eq:C_approx}
\end{align}
When the density map $\phi$ is a summation of Gaussian sources, the mean and autocorrelation features have closed form expressions, i.e. they become noncentral $\chi$ distributions with three degrees of freedom. In fact, it can be shown that the analytical expressions of the features for Gaussian sources are functions of the radial and pairwise distances of said Gaussian sources. However, due to limited space, we do not further elaborate upon their analytical forms. 

\subsection{Reconstructing 3D map}
We divide the compact support of the density map in 3D into $N=(2M+1)^3$ voxels $\{\vo_1,\cdots, \vo_N\}$, following the corresponding discretization of the projection image in \eqref{eq:discretization_image}. Let $\boldsymbol{\phi} \in \left({\mathbb{R}^+\cup\{0\}}\right)^N$ denote the nonnegative density values of the $N$ voxels. The approximated mean feature $\widetilde{\mu}(t)$ in~\eqref{eq:mu_definition} and the autocorrelation feature $\widetilde{C}(t)$ in \eqref{eq:c_definition} can be written in the following discrete forms accordingly,
\begin{align}
    \widetilde{\mu}(t) &= \vg_t^T\boldsymbol\phi\\
    \widetilde{C}(t) &= \boldsymbol\phi^T\vE_t\boldsymbol\phi\,,
\end{align}
where $\vg_t\in\{0,1\}^N$ is the measurement vector that produces the mean feature $\mu(t)$, and $\vE_t\in\{0,1\}^{N\times N}$ is the measurement matrix that produces the autocorrelation feature $C(t)$,
\begin{align}
    \vg_t[i]&=\left\{
    \begin{array}{l}
        1  \\
        0 
    \end{array}
    \begin{array}{l}
        \textnormal{if }\|\vo_i\|_2=t  \\
        \textnormal{otherwise}
    \end{array}
    \right.\\
    \vE_t[i,j]&=\left\{
    \begin{array}{l}
    1\\
    0
    \end{array}
    \begin{array}{l}
        \textnormal{if }\|\vo_i-\vo_j\|_2=t  \\
        \textnormal{otherwise.} 
    \end{array}
    \right.
\end{align}
We then reconstruct the 3D density map $\boldsymbol\phi$ using the approach proposed in \cite{Huang2018} to solve the following nonconvex optimization problem,
\vspace{-0.4cm}
\begin{align}
    \label{eq:udgp_formulation}
    \min_{\boldsymbol\phi}\quad&\sum\limits_{i=1}^{T_C} \left(\widetilde{C}(t_i)-\boldsymbol\phi^T\vE_{t_i}\boldsymbol\phi\right)^2\\
    \label{eq:radial_constraints}
    \textnormal{subject to}\quad&\widetilde{\mu}(t_j)=\ g_{t_j}^T\boldsymbol\phi,\ \forall j\in\{1,\cdots,T_\mu\}, 
\end{align}
where $T_C$ and $T_\mu$ are the number of correlation and mean features respectively. 
As detailed in \cite{Huang2018}, using a redesigned spectral initializer, the projected gradient descent method can be used to recover a solution $\boldsymbol\phi$ subject to the set of linear constraints imposed by the mean features in \eqref{eq:radial_constraints}. 
By using a denoised image $s$ as a reference, we can further reduce the set of possible voxels the density map occupies, thus reducing the search space of the reconstruction process.

\section{Numerical results}
\vspace{-0.1cm}
\label{sec:results}

\begin{figure}
\captionsetup[subfloat]{farskip=2pt,captionskip=1pt}
\centering     
\subfigure[$\textrm{RMSD} = 3.9069 $]{\label{fig:recona}\includegraphics[width=40mm]{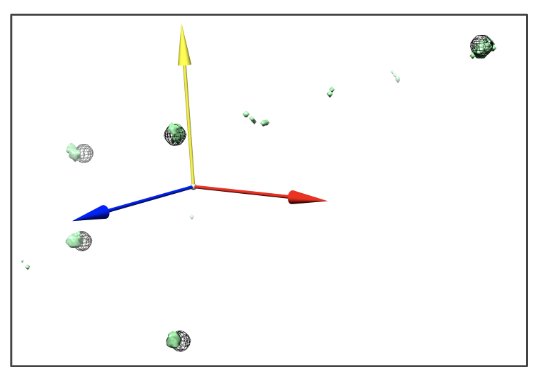}}
\subfigure[$\textrm{RMSD} = 1.0000 $]{\label{fig:reconb}\includegraphics[width=40.mm]{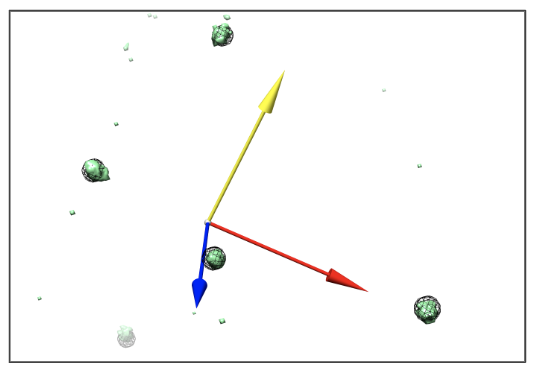}}
\subfigure[$\textrm{RMSD} = 3.6016 $]{\label{fig:reconc}\includegraphics[width=40mm]{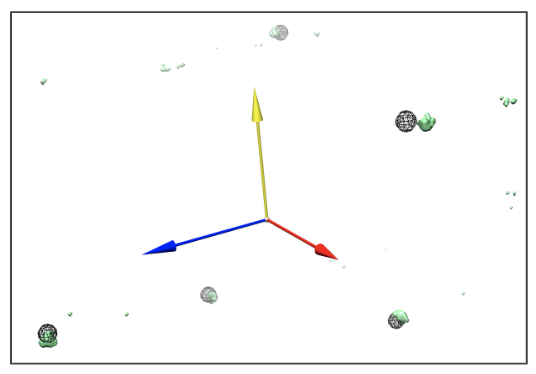}}
\subfigure[$\textrm{RMSD} = 55.7578 $]{\label{fig:recond}\includegraphics[width=40.mm]{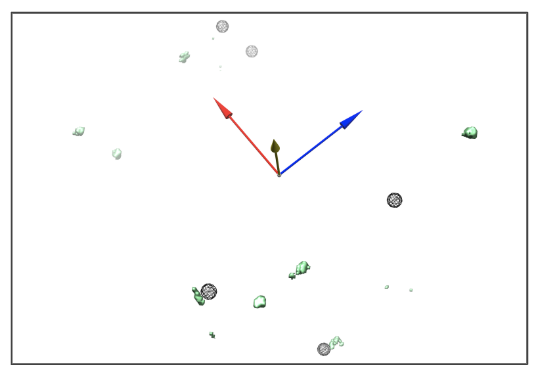}}
\caption{ A comparison between the reconstructed and ground truth 3D maps for four randomly generated point-source models. The reconstruction results of our pipeline are illustrated as green blobs and the original density map is depicted by black meshes. The arrows mark the X-Y-Z axes as red, yellow, blue. In each figure, the coordinate system is rotated accordingly for better visualization. The density maps are rendered using Chimera package~\cite{Chimera2004}. The parameters of this experiment are $L = 3\times10^4$, $\log_{10}\mathrm{SNR} = -12\, \mathrm{dB}$, $\Delta = 0.005$, $N_k=N_{\varphi}=400$.}
\label{fig:recon_results}
\end{figure}

To test the algorithm performance, we generate the coordinates of $K=5$ points randomly in the 3D volume $\left[-0.5, 0.5 \right]\times \left[-0.5, 0.5\right]\times\left[-0.5, 0.5\right]$. We set $\psi$ to be a Gaussian blob centered at the origin. 
We generate $L$ projection images following the forward model in~\eqref{eq:forward_model}, with projection views obtained by sampling $L$ points uniformly from SO(3).
From the projection images, we then compute the rotation-invariant features  using~\eqref{eq:mu_approx},\eqref{eq:C_approx}. Also, we define signal-to-noise ratio (SNR) of the projection data as the average power of the clean projection image divided by the noise power. We set the center of the mass for each blob in the reconstructed density map as the recovered location of the corresponding point-source. Because the features introduced in~\eqref{eq:mu_approx},\eqref{eq:C_approx} are rotationally invariant, the reconstructed density map is determined up to a rotation. For visualization and quantifying the error, we align the reconstruction with the ground truth and compute the root mean squared distance (RMSD) between the estimated point source locations and the ground truth (see Fig.~\ref{fig:recon_results}). 

\subsection{Ablation study}
Obtaining high quality estimations of the features is an important step towards the successful point-source reconstruction. Here we study the effect of various parameters involved in the estimation of the features. To assess the quality of the features, we rely on the $\ell_2$ distance between the estimated features derived in~\eqref{eq:mu_approx},\eqref{eq:C_approx} with their analytical expressions. Figure~\ref{fig:estimate_features} presents the quality of the estimated features with respect to the parameters including, $\Delta$ (the sampling step), $\sigma$ (the noise standard deviation), $(N_k, N_{\varphi})$ the number of discretizations over $k$ and $\varphi$ to compute the polar FFT of projection images. Figures~\ref{fig:proj}-\ref{fig:fp} demonstrate that in order to have accurately estimated features in \eqref{eq:mu_approx},\eqref{eq:C_approx} that are close to their ground truth values, we need more projection images, small-enough pixel size (i.e. sampling step $\Delta$) and sufficiently fine discretization in the Fourier domain. In addition, for a fixed $L$, higher noise regimes lead to more deviation of the estimated features from the ground truth (Fig. \ref{fig:sigma}).

\subsection{3D Gaussian source reconstruction}
Figure~\ref{fig:recon_results} showcases the reconstruction results (green blobs) in comparison to the ground truth density map (black meshes), alongside the evaluated RMSD for each reconstruction. After alignment, a reconstruction is successful if the point sources recovered by our pipeline and the ground truth are close and the final RMSD is smaller than a threshold of $10$. Figures~\ref{fig:recona}-\ref{fig:reconc} are examples of successful reconstruction where the reconstructed map overlaps with the ground truth map, leading to small RMSD. On the other hand, Fig.~\ref{fig:recond} shows an example of a failed reconstruction with a large RMSD. We ran our pipeline for $50$ randomly generated point-source models, and $72\%$ of them were successfully reconstructed. As the problem in~\eqref{eq:udgp_formulation} is nonconvex, a failed reconstruction may occur when the solution gets stuck in a local optimum far from the ground truth.

\begin{figure}
\centering    
\subfigure[]{\label{fig:noisy_data}\includegraphics[width=0.52\columnwidth]{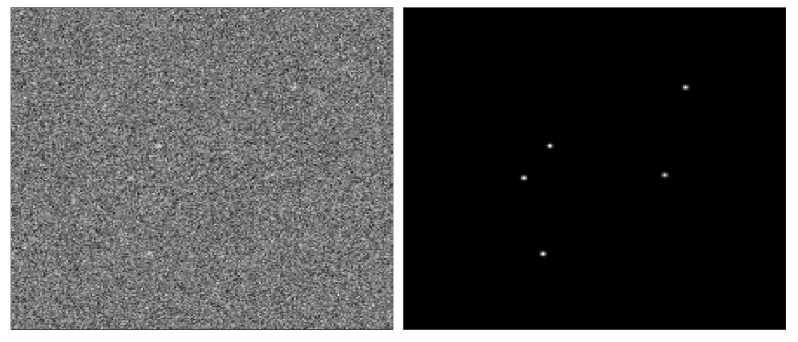}}
\subfigure[]{\label{fig:noisy_recon}\includegraphics[width=0.45\columnwidth]{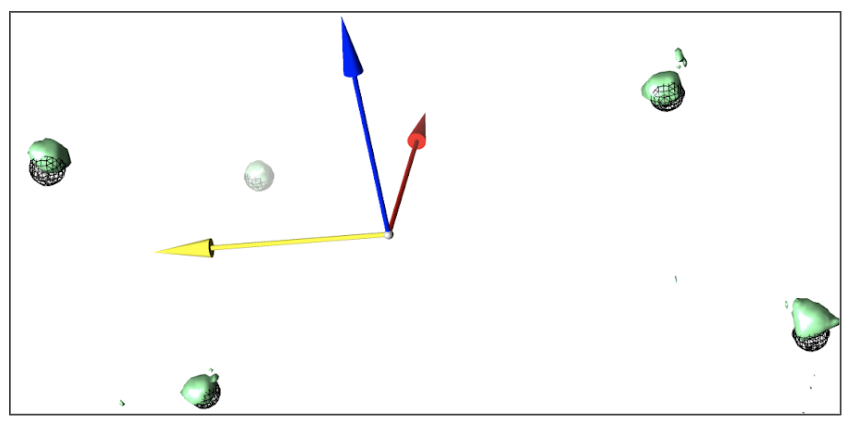}}
\vspace{-0.2cm} \\
\subfigure[]{\label{fig:noisy_feat}\includegraphics[width=1\columnwidth]{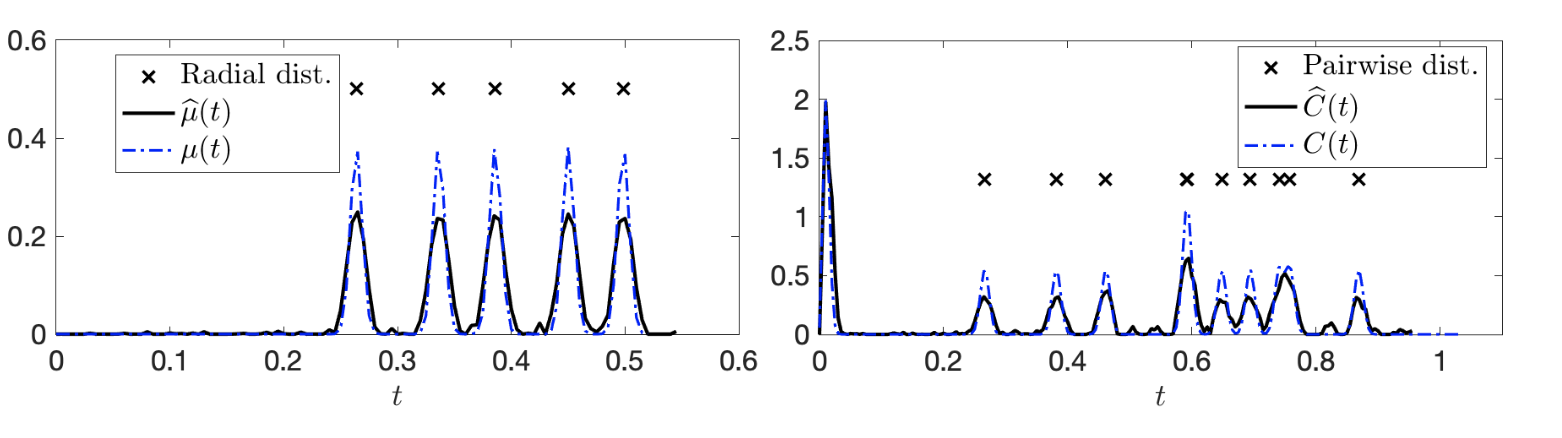}}
\vspace{-0.2cm}
\caption{Results of the experiment in low SNR regime ($\log_{10}\mathrm{SNR} = -24.83\, \textrm{dB}$). \protect \subref{fig:noisy_data} An example of the noisy projection image (left) and its clean version (right). \protect \subref{fig:noisy_recon} Comparison between the 3D reconstructed (green blob) and ground truth density map (black mesh). \protect\subref{fig:noisy_feat} Comparison between the estimated mean and auto-correlation features from the projection data (black curves), the groundtruth expressions of the features (blue dashed curves), the radial and pairwise distances of the Gaussian point source model.}
\label{fig:noisy_exp}
\vspace{-0.5cm}
\end{figure}

\subsection{Reconstruction in a low SNR regime}
We further tested the robustness of our pipeline in a significantly low $\mathrm{SNR}$ regime ($\log_{10}\mathrm{SNR} = -24.83\, \textrm{dB}$). In Fig.~\ref{fig:noisy_data} we show one example of the noisy projection images used to estimate the features. The clean projection image is also provided for reference. In order to estimate the features, we ended up using $L=10^6$ noisy projection images. 
The estimated mean and autocorrrelation features are displayed in Fig.~\ref{fig:noisy_feat}. Note that, as expected, the peaks of the mean and autocorrelation features coincide with the radial and pairwise distances of the Gaussian point source model. In addition, we see that the estimated features (solid black curves) closely resemble the ground truth (blue dashed curves). Finally, Fig.~\ref{fig:noisy_recon} compares the 3D reconstructed model (green blobs) with the ground truth density map (black meshes), confirming the successful reconstruction of the point source model from the features with $\textrm{RMSD}=1.9$.

\section{Conclusion}
\vspace{-0.1cm}
\label{sec:conclusion}
In  this  paper,  we  considered  the  problem  of  reconstructing  a  3D point-source  model  from  a  set  of  2D  noisy  projection  data  taken from unknown view angles.  Compared to conventional approaches that iteratively estimate the view angles, our proposed approach relies on mean and autocorrelation features computed from a large collection of projection images. The reconstruction problem is then formulated as a constrained nonconvex problem, which can be solved using the projected gradient descent with a spectral initialization strategy.  Numerical experiments show the potential of our approach in extracting robust interpretable features and operating in extremely low SNR settings. In the future work we would like to explore how to incorporate other rotation invariant features and apply our method on real projection data with general 3D density map.

\newpage
\bibliographystyle{IEEEbib}
\bibliography{references.bib}

\end{document}